\documentclass[twocolumn,showpacs,preprintnumbers,amsmath,amssymb]{revtex4-1}
\usepackage{graphicx}% Include figure files
\usepackage{bm}% bold math

\begin{document}

\preprint{APS/123-QED}

\title{Self-heating in kinematically complex magnetohydrodynamic flows}% Force line breaks with \\
\author{Zaza Osmanov}
%\email{z.osmanov@astro-ge.org}
\altaffiliation[Also at ] {Free University, Bedia str. 0183-Tbilisi,
Georgia}
 \affiliation{Centre for Theoretical Astrophysics, ITP, Ilia State University, 0162-Tbilisi, Georgia
}%Lines break automatically or can be forced with \\

\author{Andria Rogava}%
 %\email{a.edu}
\affiliation{Center for Theoretical Astrophysics, ITP, Ilia State
University, 0162-Tbilisi, Georgia
}%

\author{Stefaan Poedts}
\affiliation{Centre for Plasma Astrophysics, Katholieke Universiteit
Leuven, Celestijnenlaan 200B, bus 2400 B-3001, Belgium% with \\
}%

\date{\today}% It is always \today, today,

\begin{abstract}
The non-modal self-heating mechanism driven by the velocity shear in
kinematically complex magnetohydrodynamic (MHD) plasma flows is
considered. The study is based on the full set of MHD equations
including dissipative terms. The equations are linearized and
unstable modes in the flow are looked for. Two different cases are
specified and studied: (a) the instability related to an exponential
evolution of the wave vector; and (b) the parametric instability,
which takes place when the components of the wave vector evolve in
time periodically. By examining the dissipative terms, it is shown
that the self-heating rate provided by viscous damping is of the
same order of magnitude as that due to the magnetic resistivity. It
is found that the heating efficiency of the exponential instability
is higher than that of the parametric instability.

\end{abstract}

\maketitle

%%%%%%%%%%%%%%%%%%%%%%%%%%%%%%%%%%%%%%
\section{Introduction}
%%%%%%%%%%%%%%%%%%%%%%%%%%%%%%%%%%%%%%

It is widely known that astrophysical, terrestrial and laboratory
plasma flows are most often characterized by spatially inhomogeneous
velocities, constituting so-called shear flows. In many cases, these
flows are kinematically complex, multidimensional and sometimes they
are also relativistic. Among the different kinds of plasma shear
flows many exhibit motion profiles with highly nontrivial velocity
fields. One prominent class of such flows contains the swirling
astrophysical flows that until now have received only little
theoretical attention. In particular, many galaxies reveal
relativistic jets \cite{tavec}, some of which are characterized by
helical motion patterns \cite{broder,kharb}. A similar situation is
met in pulsars, where morphological and spectroscopic studies make
strong jet-like structures evident \cite{crabjet,pjet1}. Even closer
to us, namely in the solar atmosphere, the existence of giant
macro-spicules featuring swirling, tornado-like plasma motions has
been reported \cite{pm98}.

The latter class of jet-like large-scale velocity patterns in the
atmosphere of the Sun deserves some special attention. In 1991, the
{\it Yohkoh} Solar Observatory detected coronal jets (by means of
the Soft X-Ray Telescope \cite{tsun}) which were assumed to be
related to evaporation flows \cite{shimo}. More recently, it
%%%??? what do you mean by "it" here???
has been reported on the $^3$He-rich solar energetic particle event
\cite{sjet} observed on 2006 November $18\,$th by the X-Ray
Telescope on board the Hinode satellite \cite{golub}. There is also
some indirect observational evidence for the presence of more
complicated velocity patterns. In particular, blue shifted and red
shifted emission on either side of the axis of a macro-spicule has
been reported \cite{pm98}. These observations were interpreted as
indicating at the presence of a rotation within these tall pillars
of hot and magnetized, swirling plasma flows. Another interesting
class of astronomical objects with an observational evidence of
swirling flows is the Herbig-Haro jets. It was shown that the young
stellar object HH~212 exhibits strong indications for the presence
of rotation in the jet \cite{davis}. A similar helical structure is
observed in DG~Tau \cite{bacc}.

Therefore, the study of kinematically complex shear flows with a
non-trivial geometry is interesting and necessary because flow
patterns with nontrivial velocity fields appear to be quite common
in astrophysics. Besides, it seems reasonable to expect that the
kinematic complexity could considerably influence the physical
processes within these flows and lead to various detectable
observational appearances in the related astronomical objects.

Recently, it was fully realized that collective phenomena in shear
flows are characterized by so-called non-modal processes, related
with non-normality \cite{tref}. These processes are fairly
well-understood for a variety of flows, see, for instance,
\cite{rmb97,bod01b,andro,chven}. In particular, in
\cite{andro,chven} shear flow instabilities are studied in helical
flows. The problem was considered both for the incompressible
\cite{andro} and compressible \cite{chven} cases and it has been
shown that shear flows efficiently amplify the fast/slow
magnetosonic waves (in the case of compressible flows) and the
Alfv\'en waves (in both cases). It was also argued that these
instabilities may lead to  so-called 'non-modal self-heating'. This
process involves three major stages: (a) First, the waves are
excited spontaneously within the shear flow; (b) second, the excited
modes undergo a non-modal amplification, extracting a part of the
equilibrium flow's kinetic energy and (c) third, in the final stage,
the non-modally amplified waves get dissipated via viscous damping
and/or magnetic resistive diffusion and give their energy back to
the background flow in the form of heat.

This mechanism was originally described in \cite{androSH} for the
case of a hydrodynamic flow and it has been argued that self-heating
may play an important role also in the dynamics of kinematically
complex magnetized flows. In \cite{lcl06} it has been shown that
efficient non-modal self-heating can indeed take place in magnetized
media as well. A direct application to the solar atmosphere was
proposed in \cite{spp06} where  it was concluded that self-heating
could serve as an efficient mechanism for providing, or at least
contributing to, the heat production in the lower corona. Recently,
we have studied the efficiency of non-modal self-heating by acoustic
wave perturbations in the solar chromosphere \cite{rop10}. It has
been shown that amplified waves get damped due to the presence of
viscous dissipation and indeed cause an energy transfer back to the
background flow in the form of heat.

In most of the above cited papers, the problem of self-heating was
considered only for shear flows with relatively simple flow
geometry, while the flows occurring in astrophysical plasmas, as
well as in many terrestrial and laboratory situations, can be
significantly more complex. Therefore, it is quite timely and
important to examine kinematically more complicated cases and verify
if the mechanism is also efficient when such complex flow patterns
occur. For this purpose, in the present paper we study the
possibility of non-modal self-heating in fully compressible,
kinematically complex MHD flows taking into account both viscosity
and magnetic resistivity.

The paper is arranged in the following way. In section~II, we
formulate the model for the study of non-modal shear flow
instabilities for MHD flows including resistive terms and we derive
the equations governing the self-heating process. In section~III, we
apply our model to the fully compressible flows examining the high
plasma-$\beta$ case and we then investigate numerically the
efficiency of the self-heating mechanism. In section~IV, we discuss
the obtained results and outline the directions of further, more
detailed, quantitative and applicative studies.

%%%%%%%%%%%%%%%%%%%%%%%%%%%%%%%%%%%%%%%%%%%
\section[]{Main Consideration}
%%%%%%%%%%%%%%%%%%%%%%%%%%%%%%%%%%%%%%%%%%%

In order to study the self-heating mechanism in magnetized plasma
flows, we start with the standard set of polytropic MHD equations
consisting of the equation of mass conservation:
\begin{equation}
\label{conn} D_t\rho + \rho \nabla \cdot {\bf  V}= 0,
\end{equation}
the momentum conservation equation:
\begin{equation}
\label{euler} D_t{\bf V} = -\frac{1}{\rho}{\bf \nabla P} -\frac{{\bf
B}}{4\pi\rho}\times ({\bf \nabla\times B}) + \nu\Delta{\bf V},
\end{equation}
the induction equation:
\begin{equation}
\label{indd} D_t{\bf B} = ({\bf B} \cdot \nabla){\bf V} -{\bf B}(\nabla \cdot {\bf V})
+\eta\Delta{\bf B},
\end{equation}
Gauss's law for magnetism:
\begin{equation}
\label{divbb} {\bf \nabla\cdot B}=0,
\end{equation}
and the polytropic equation of state:
\begin{equation}
\label{eqstate} P=C{\rho}^{1+1/n},
\end{equation}
where $D_t \equiv \partial_t + ({\bf V} \cdot \nabla)$ denotes the
convective derivative, $\rho$ is the density, $P$ stands for the
pressure, ${\bf V}$ is the velocity field, ${\bf B}$ denotes the
magnetic field, $\nu$ stands for the coefficient of kinematic
viscosity, $\eta$ is the magnetic resistivity coefficient, and
$C={\it const}$ and $n$ are the polytropic constant and the
polytropic index, respectively. Note that in the framework of the
present paper we consider a common stellar model, viz.\ the
so-called ``Polytropic Star Model'' in which the pressure depends
only on the density according to the polytropic equation of state,
Eq.~(5) \cite{carroll}.

The equilibrium state of the model is specified by a homogeneous MHD
plasma ($\rho_0=\hbox{\it const}$), embedded in a uniform, vertical
magnetic field (${\bf B_0}\equiv [0, 0, B_0=\hbox{\it const}]$). For
analyzing the full set of MHD equations, we first linearize them,
perturbing all physical quantities about the equilibrium state:
\begin{equation}
\label{dec1} \rho \equiv \rho_0 + \rho',
\end{equation}
\begin{equation}
\label{dec2} {\bf V} \equiv {\bf U} + {\bf v},
\end{equation}
\begin{equation}
\label{dec3} {\bf B} \equiv {\bf B_0} + {\bf B'},
\end{equation}
and
\begin{equation}
\label{dec4} P \equiv P_0 + P'.
\end{equation}
It is assumed that the perturbations $\rho'$, ${\bf v}$, ${\bf B'}$
and $P'$ are small compared to the corresponding equilibrium
quantities, viz.\ $\rho_0$, ${\bf U}$, ${\bf B_0}$ and $P_0$,
respectively.
By substituting Eqs.~(\ref{dec1}-\ref{dec4}) into
Eqs.~(\ref{conn}-\ref{eqstate}), and then linearizing (i.e.\ keeping
only the first-order terms in the perturbed quantities), one can
derive the following set of equations for the perturbed quantities:
\begin{equation}\label{con}
\mathcal{D}_tD + {\bf \nabla} \cdot {\bf v}= 0,
\end{equation}
\begin{equation}\label{eul}
\mathcal{D}_t{\bf v} + ({\bf v} \cdot \nabla){\bf U} = -C_s^2{\bf
\nabla} D+ C_A^2[\partial_z{\bf b}-{\bf\nabla}b_z] + \nu\Delta{\bf
v},\end{equation}
\begin{equation}
\label{ind} \mathcal{D}_t{\bf b} = ({\bf b\cdot\nabla )U}+
\partial_z{\bf v}+{\bf e_z(\nabla\cdot v)}+\eta{\bf \Delta b},
\end{equation}
\begin{equation}
\label{divb} {\bf \nabla\cdot b}=0,
\end{equation}
where $\mathcal{D}_t \equiv \partial_t + ({\bf U}\cdot \nabla)$,
$D\equiv\rho'/\rho_0$, ${\bf b}\equiv{\bf B'}/B_0$, $C_s =\sqrt{
dP_0/d\rho_0}$ denotes the sound speed and $C_A\equiv
B_0/\sqrt{4\pi\rho_0}$ is the Alfv\'en speed.

Following the method developed in \cite{mahand}, we assume that
${\bf U}$ is a spatially inhomogeneous vector velocity field. We
expand the velocity in Taylor series around a point
$A(x_0,y_0,z_0)$, preserving only the linear terms:
\begin{equation}\label{velexpand}
{\bf U}={\bf U}(A)+\sum_{i=1}^3\frac{\partial{\bf U}(A)}{\partial
x_i}(x_i-x_{i0}),\end{equation}
where $i=1,2,3$ and $x_i=(x,y,z)$.

Mahajan and Rogava have shown that by applying the following ansatz
\begin{equation}\label{anzatz}
F(x,y,z,t)\equiv\hat{F}(t)e^{\phi_1-\phi_2},\end{equation}
\begin{equation}\label{fi1}
\phi_1\equiv\sum_{i=1}^3{K_i}(t)x_i,\end{equation}
\begin{equation}\label{fi2}
\phi_2\equiv\sum_{i=1}^3U_i(A)\int{K_i}(t)dt,\end{equation}
to Eqs.~(\ref{con}-\ref{eul}), these equations transform to a set of
{\em ordinary} differential equations \cite{mahand}. This reduces
the problem mathematically to the study of an initial value problem.
By $U_i(A)$ we denote the equilibrium velocity components and the
$K_i(t)$ represent the wave number vector components satisfying the
following set of differential equations \cite{mahand}:
\begin{equation}\label{dk}
{\bf \partial_{t}K}+ {\bf S^T} \cdot {\bf K}=0,\end{equation}
where ${\bf S^T}$ is the transposed shear matrix. This shear matrix
${\bf S}$ is defined by:
\begin{equation}\label{S}
 {\bf S} = \left(\begin{array}{ccc} U_{x,x} & U_{x,y} & U_{x,z}  \\
U_{y,x} & U_{y,y} & U_{y,z}  \\ U_{z,x} & U_{z,y} & U_{z,z} \\
\end{array} \right ),\end{equation}
with $U_{i,k}\equiv\partial U_{i}/\partial x_k$.

Working with the background velocity field ${\bf
U(r)}\equiv[0,r\Omega(r),U(r)]$, the shear matrix obtains the
following form \cite{andro,chven}
\begin{equation}\label{S1}
 {\bf S} = \left(\begin{array}{ccc} \Sigma & A_1 & 0  \\
A_2 & -\Sigma & 0  \\ C_1 & C_2 & 0 \\
\end{array} \right ).\end{equation}

% Here we need to specify new quantitites introduced in Eq.(20) !!!

By taking Eqs.~(\ref{anzatz}-\ref{dk},\ref{S1}) into account, the
set of Eqs.~(\ref{con}-\ref{divb}) reduces to the following
dimensionless form
\begin{equation}\label{cont}
d^{(1)}={\bf k} \cdot {\bf u},\end{equation}
\begin{equation}\label{v}
{\bf u}^{(1)}+{\bf s}\cdot{\bf u} = -{\bf k}\epsilon^2d+{\bf h}-{\bf
k}h_z-\overline{\nu}k^2 {\bf u},\end{equation}
\begin{equation}\label{h}
{\bf h}^{(1)}={\bf s\cdot h-u-e_z(k\cdot u)}-\overline{\eta}k^2 {\bf
h},\end{equation}
\begin{equation}\label{divh}
{\bf k\cdot h}=0,\end{equation}
\begin{equation}\label{k}
\textbf{k}^{(1)} + {\bf s^T} \cdot {\bf k} = 0, \end{equation}
with
\begin{equation}\label{S2}
 {\bf s} = \left(\begin{array}{ccc} \sigma & a_1 & 0  \\
a_2 & -\sigma & 0  \\ r_1 & r_2 & 0 \\
\end{array} \right ),\end{equation}
where $d \equiv iD$, ${\bf u} \equiv {\bf v}/C_A$, ${\bf h}\equiv
i{\bf b}$, $\epsilon\equiv C_s/C_A$, $a_{1,2} \equiv
A_{1,2}/K_zC_A$, $\sigma \equiv \Sigma/K_zC_A$, $r_{1,2} \equiv
C_{1,2}/K_zC_A$, $\textbf{k}\equiv \textbf{K}/K_z$,
$\overline{\nu}\equiv\nu K_z/C_A$ and $\overline{\eta}\equiv\eta
K_z/C_A$. By $F^{(1)}$ we denote $dF/d\tau$ and $\tau\equiv K_zC_At$
is the dimensionless time.

For estimating the efficiency of the self-heating mechanism, it is
useful to define the total energy of the perturbations:
\begin{equation}\label{etot}
E_{tot}\equiv E_{kin}+E_{m}+E_{c}=\frac{{\bf u^2}}{2}+\frac{{\bf
h^2}}{2}+\frac{{\epsilon^2d^2}}{2},\end{equation}
which consists of the kinetic, magnetic and compressional energies,
respectively. The evolutionary equation for the total energy has the
following form:
$$E^{(1)}_{tot}=-{\bf u(s\cdot u)}+{\bf h(s\cdot h)}-2h_z{\bf (k\cdot u)}-$$
\begin{equation}\label{e1}
\;\;\;\;\;\;\;\;\;\;\;\;\;\;\;\;\;\;\;\;\;\;\;\;\;\;\;\;\;\;\;\;\;\;\;\;\;\;\;\;
\;\;\;\;\;\;\;\;\;\;-\overline{\nu}{\bf k^2u^2}-\overline{\eta}{\bf
k^2h^2}.\end{equation}
As it is clear from Eq.~(\ref{e1}), the last two (viscosity and
magnetic resistivity) terms are responsible for the energy
dissipation and thus for the final stage of the self-heating
process. By means of these two terms one can define the so-called
'self-heating rate', originally introduced in \cite{androSH} only
for purely hydrodynamic flows:
\begin{equation}\label{psi}
\Psi(\tau) \equiv \frac{1}{E_{tot}(0)} \int_0^\tau
\left[\overline{\nu}{\bf k^2}(\tau'){\bf u^2}(\tau')
+\overline{\eta}{\bf k^2}(\tau'){\bf
h^2}(\tau')\right]d\tau'.\end{equation}

%\begin{figure}
% \par\noindent
% {\begin{minipage}{0.49\linewidth}
% \includegraphics[width=\textwidth] {fig2a.eps}
% \end{minipage}
% }
% \hfill
% {\begin{minipage}{0.49\linewidth}
% \includegraphics[width=\textwidth] {fig2b.eps}
% \end{minipage}
% }
% \caption[ ] {The temporal evolution of the self-heating rate for two different cases: on the
%left panel we examine $\overline{\nu} = 1.25\times 10^{-4}$,
%$\overline{\eta} = 0$, while on the right panel the result is shown
%for $\overline{\nu} = 0$, $\overline{\eta} = 1.25\times 10^{-4}$.
%The following set of parameters: $\sigma = 0$, $a_1 = -a_2 = -1$,
%$r_1 = 0.01$, $r_2 = 0.9406$, $\epsilon = 0.1$, $k_{x0}=k_{y0}= 10$,
%$u_{x0} = u_{y0} = u_{z0} = 0$, $h_{x0} = 0.01$, $h_{y0} = 0$, $d_0
%= 0$.}\label{fig2}
% \end{figure}

%%%%%%%%%%%%%%%%%%%%%%%%%%%%%%%%%%%%%%%%%%%%%%%%%%%%%%%%%%%%%%%%%
\section{Discussion}
%%%%%%%%%%%%%%%%%%%%%%%%%%%%%%%%%%%%%%%%%%%%%%%%%%%%%%%%%%%%%%%%%%%%%%%%%%%%

In this section, we study the efficiency of the self-heating
mechanism for different values of parameters, taking into account
both the viscous damping and magnetic resistivity terms. Generally
speaking, for the non-modal self-heating process to be efficient,
the dissipative terms must not be too small in order to provide
substantial heating.  However, at the same time the damping  should
not be ``too strong'' either: non-modally amplified modes have to
have enough time for growing and extracting the equilibrium flow
energy before damping occurs. As it is clear from
Eqs.~(\ref{euler}-\ref{indd}), the effects of dissipation become
important only for relatively small length scales.

Mathematically the non-modal instability is described by the system
of equations (\ref{cont}-\ref{k}), where Eq.~(\ref{k}) plays an
important role. In particular, in plane-parallel flows ${\bf k}$
exhibits a linear time evolution \cite{rog00}, whereas for
multidimensional cases (such as for helical flows), the components
of the wave vector may evolve in time either exponentially or
periodically \cite{andro,chven}. Linearly and exponentially evolving
wave vectors lead to so-called 'usual' instabilities (with
$\Gamma^2\equiv \sigma^2+a_1a_2>0$). Parametrically unstable modes
deserve a special interest. These modes appear for periodically
evolving ${\bf k}$'s (i.e.\ $\Gamma^2<0$) for certain values of the
parameters.
\begin{figure}
  \resizebox{\hsize}{!}{\includegraphics[angle=0]{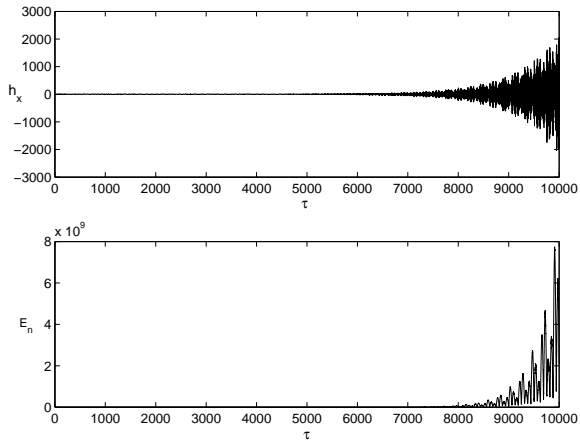}}
  \caption{The temporal behavior of $h_x(\tau)$ and the energy $E_n(\tau)\equiv
  E_{tot}(\tau)/E_{tot}(0)$ normalized by its initial value for the following
  set of parameters: $\sigma = 0$, $a_1 = -a_2 = -1$,
  $r_1 = 0.01$, $r_2 = 0.9406$, $\epsilon = 0.1$, $k_{x0}=k_{y0}=
  10$, $u_{x0} = u_{y0} = u_{z0} = 0$, $h_{x0} = 0.01$, $h_{y0} = 0$,
  $d_0 = 0$, $\overline{\nu} = \overline{\eta} = 0$. Note that the range of $r_2$,
  where the evolution of the  modes is parametrically unstable, is very
  narrow: $[0.94; 0.941]$.}\label{fig1}
\end{figure}
In Fig.~\ref{fig1}, we show the temporal behavior of $h_x(\tau)$ and
the energy $E_n(\tau)\equiv E_{tot}(\tau)/E_{tot}(0)$ normalized by
its initial value. The considered set of parameters is: $\sigma =
0$, $a_1 = -a_2 = -1$, $r_1 = 0.01$, $r_2 = 0.9406$, $\epsilon =
0.1$, $k_{x0}=k_{y0}= 10$, $u_{x0} = u_{y0} = u_{z0} = 0$, $h_{x0} =
0.01$, $h_{y0} = 0$, $d_0 = 0$, $\overline{\nu} = \overline{\eta} =
0$. As it is clear from the plots, despite $\Gamma^2<0$, the system
exhibits unstable modes. The instability takes place only for very
narrow ranges of parameters. In particular, one can
straightforwardly check that the instability disappears for values
of $r_2$ outside the interval $[0.94; 0.941]$.
\begin{figure}
\resizebox{\hsize}{!}{\includegraphics{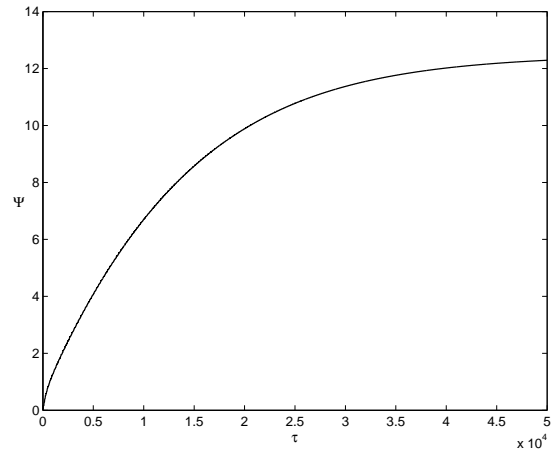}}
\resizebox{\hsize}{!}{\includegraphics{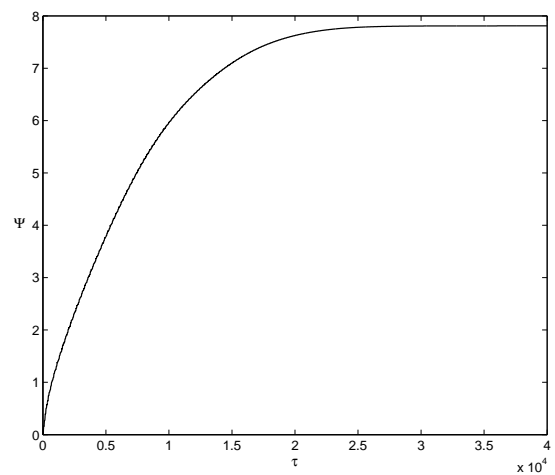}}
 \caption{The temporal evolution of the self-heating rate for two different cases.
 {\it Top panel}: $\overline{\nu} = 1.25\times 10^{-4}$,
$\overline{\eta} = 0$.  {\it Bottom panel}: $\overline{\nu} = 0$,
$\overline{\eta} = 1.25\times 10^{-4}$. The set of parameters is
$\sigma = 0$, $a_1 = -a_2 = -1$, $r_1 = 0.01$, $r_2 = 0.9406$,
$\epsilon = 0.1$, $k_{x0}=k_{y0}= 10$, $u_{x0} = u_{y0} = u_{z0} =
0$, $h_{x0} = 0.01$, $h_{y0} = 0$, $d_0 = 0$.}
 \label{fig2}
\end{figure}

The dissipative factors will inevitably lead to the self-heating
effect. In particular, Fig.~\ref{fig2} shows the temporal evolution
of the self-heating rate for two different cases: on the top panel
we examine $\overline{\nu} = 1.25\times 10^{-4}$, $\overline{\eta} =
0$ and on the bottom panel the result is shown for $\overline{\nu} =
0$ and $\overline{\eta} = 1.25\times 10^{-4}$. All the other
parameters are the same as in Fig.~\ref{fig1}. As we see from the
plots, there is no principal difference between viscosity-dominated
and magnetic-resistivity-dominated cases, i.e. the saturated values
of the self-heating rates are of the same order of magnitudes. In
both cases, the energy converted into heat exceeds the initial
perturbation energy about 10 times. This means that the  excited
non-modal waves extract energy from the mean flow and this energy,
via the agency of the dissipative terms, is converted into heat and given
back to the flow. As we see from the results, the efficiency of the
self-heating is quite high.

%\begin{figure}
% \par\noindent
% {\begin{minipage}{0.5\linewidth}
% \includegraphics[width=\textwidth] {fig3a.eps}
% \end{minipage}
% }
% \hfill
% {\begin{minipage}{0.44\linewidth}
% \includegraphics[width=\textwidth] {fig3b.eps}
% \end{minipage}
% }
% \caption[ ] {The temporal evolution of the self-heating rate for three different cases: on the
%left panel we show the results for $\overline{\nu} = 1.2\times
%10^{-5}$ (solid line) and $\overline{\nu} = 1.24\times 10^{-5}$
%(dashed line) and on the right panel the result is shown for
%$\overline{\nu} = 1.22\times 10^{-5}$. The set of parameters is
%$\sigma = 0$, $a_1 = -a_2 = -1$, $r_1 = 0.01$, $r_2 = 0.9406$,
%$\epsilon = 0.1$, $k_{x0}=k_{y0}= 10$, $u_{x0} = u_{y0} = u_{z0} =
%0$, $h_{x0} = 0.01$, $h_{y0} = 0$, $d_0 = 0$.}\label{fig3}
% \end{figure}

\begin{figure}
\resizebox{\hsize}{!}{\includegraphics{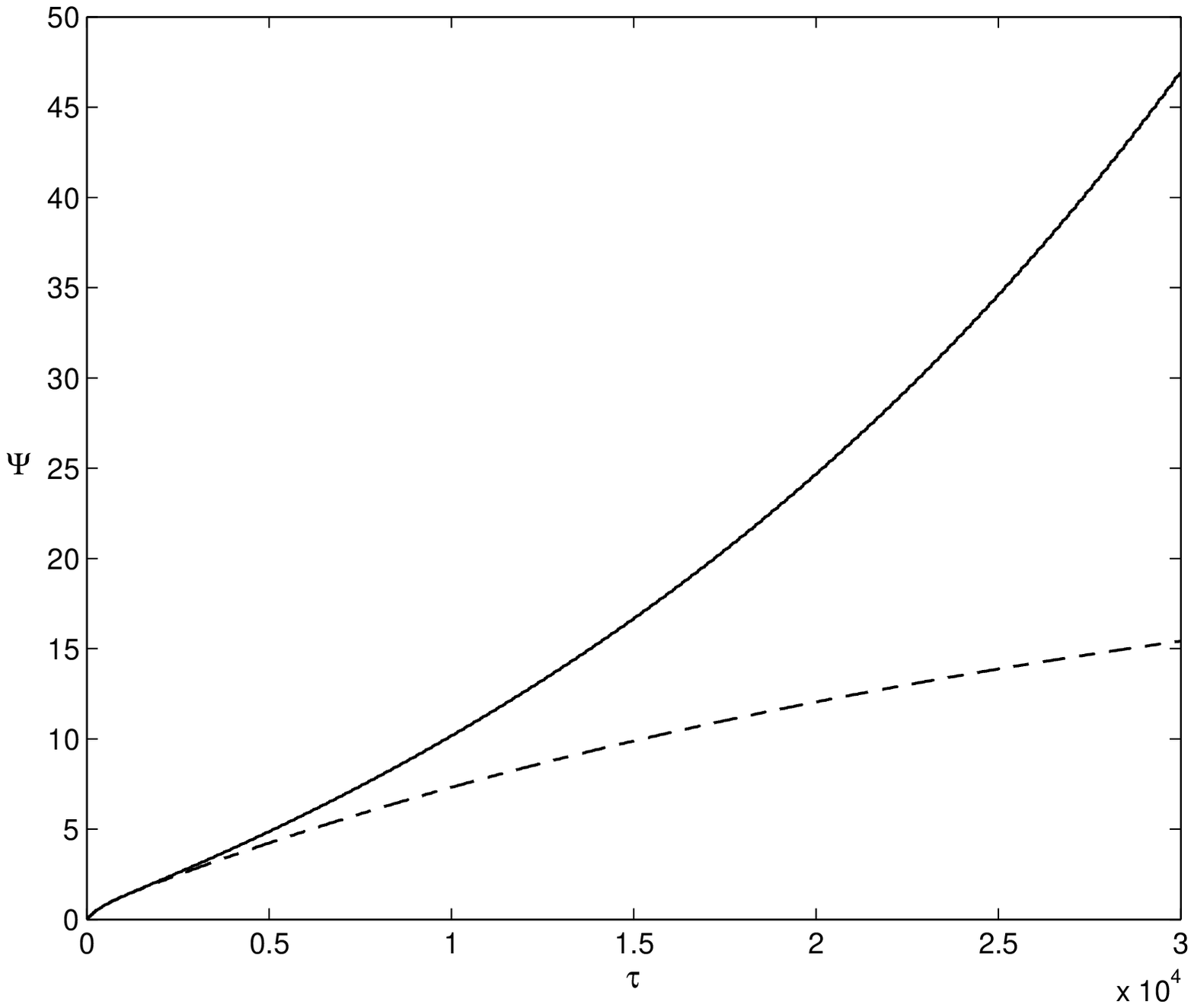}}
\resizebox{\hsize}{!}{\includegraphics{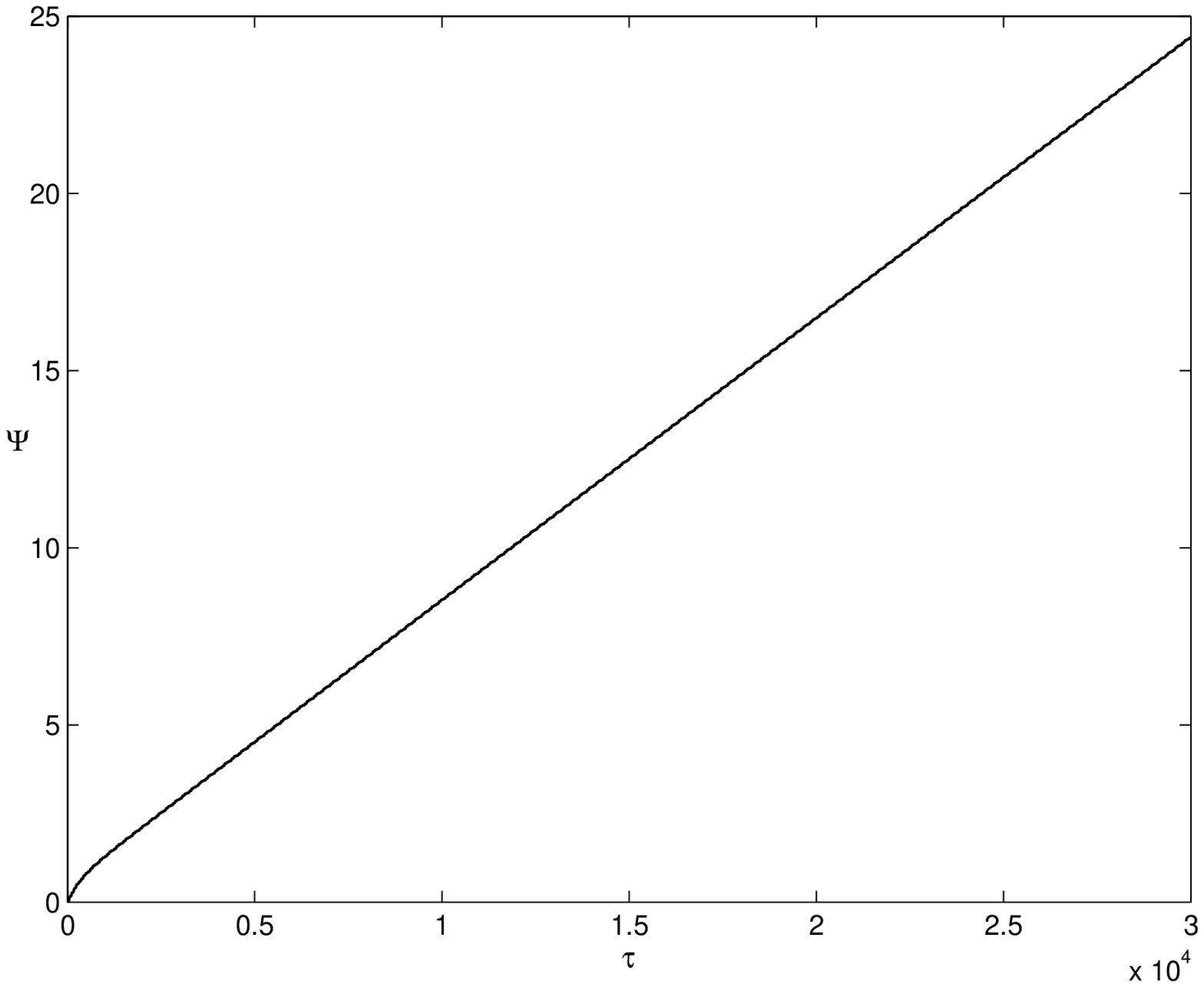}}
 \caption{The temporal evolution of the self-heating rate for three different cases.
 {\it Top panel}: $\overline{\nu} = 1.2\times
10^{-5}$ (solid line) and $\overline{\nu} = 1.24\times 10^{-5}$
(dashed line). {\it Bottom panel}: $\overline{\nu} = 1.22\times
10^{-5}$. The set of parameters is $\sigma = 0$, $a_1 = -a_2 = -1$,
$r_1 = 0.01$, $r_2 = 0.9406$, $\epsilon = 0.1$, $k_{x0}=k_{y0}= 10$,
$u_{x0} = u_{y0} = u_{z0} = 0$, $h_{x0} = 0.01$, $h_{y0} = 0$, $d_0
= 0$.}
 \label{fig3}
\end{figure}
\begin{figure}
  \resizebox{\hsize}{!}{\includegraphics[angle=0]{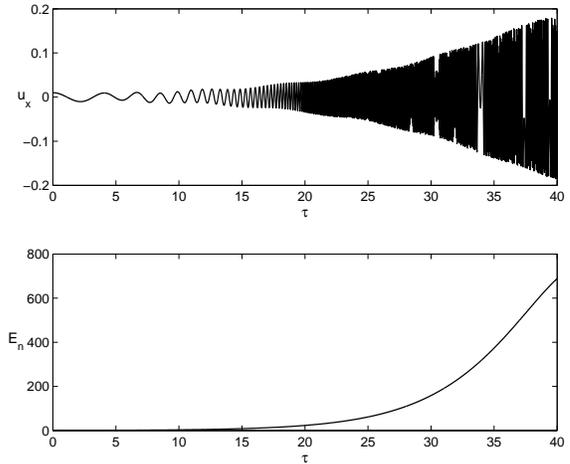}}
  \caption{The temporal behavior of $u_x(\tau)$ and $E_n(\tau)$.
  The considered set of parameters is $\sigma = 0$, $a_1 = a_2 = 1$,
  $r_1 = r_2 = 0$, $\epsilon = 0.1$, $k_{x0}= 1$, $k_{y0}=
  0$, $u_{x0} = 0$, $u_{y0} = u_{z0} = 0$, $h_{x0} = h_{y0} = 0$,
  $d_0 = 0$, $\overline{\nu} = \overline{\eta} = 0$.}\label{fig4}
\end{figure}

It is worth noting that for the considered mechanism to be
efficient, several conditions must be satisfied. At the one hand,
the instability must be robust enough to provide an amplification of
the waves and, additionally, the dissipation process must not be too
efficient in order to avoid damping of the excited modes before they
grow substantially enough. On the other hand, if the damping is too
weak the amplified waves fail to give back their energy back to the
flow. Therefore, one may expect that self-heating will be efficient
for a limited range of moderatly dissipative systems.

On the top panel of Fig.~\ref{fig3}, we show the temporal evolution
of the self-heating rate for $\overline{\nu} = 1.2\times 10^{-5}$
(solid line) and $\overline{\nu} = 1.24\times 10^{-5}$ (dashed
line). The other parameters are again the same as in
Fig.~\ref{fig1}. As it is evident from the plots, for the case
presented by the solid line, the viscosity is not strong enough and
non-modaly amplified modes avoid any significant viscous
dissipation. In general, the temporal evolution of the self-heating
rate is very sensitive to viscosity. By slightly increasing the
value of the viscosity, the situation might change drastically. By
the dashed line we show the result for $\overline{\nu} = 1.24\times
10^{-5}$, and it is clear that unlike the previous case, more
dominant viscous terms here ensure stronger self-heating. The
efficiency of the instability seems significantly decreased and
$\Psi(t)$ tends to become saturated (see Fig.~\ref{fig2}). This
tendency implies that for a certain value of the viscosity the
effects of non-modal instability and dissipation will balance each
other. This particular case is presented on the bottom panel of
Fig.~\ref{fig3}, where the value of the viscosity equals $1.22\times
10^{-5}$ and, as we see from the figure, the self-heating rate shows
a linear dependence on time.

We have already noticed that apart from the parametric instability,
the system under consideration exhibits also `usual' instabilities
($\Gamma^2>0$), characterized by exponentially evolving wave number
vector. Normally, in these cases the waves are more unstable than
the parametrically unstable modes. In particular, in Fig.~\ref{fig4}
we show the temporal behavior of $u_x(\tau)$ and $E_n(\tau)$. The
considered set of parameters is: $\sigma = 0$, $a_1 = a_2 = 1$, $r_1
= r_2 = 0$, $\epsilon = 0.1$, $k_{x0}= 1$, $k_{y0}= 0$, $u_{x0} =
0$, $u_{y0} = u_{z0} = 0$, $h_{x0} = h_{y0} = 0$, $d_0 = 0$,
$\overline{\nu} = \overline{\eta} = 0$. As we see from the figure,
the energy perturbation and $u_x$ significantly increase until $\tau
= 40$. These results can be explained as follows: since
$\Gamma^2>0$, the wave vector components evolve with time
exponentially (unlike the cases presented by figures
\ref{fig1}-\ref{fig3}) and correspondingly, the system undergoes a
strong instability, the growth rate of which exceeds that of the
parametric instability. For the same reason, the efficiency of the
self-heating mechanism for the unstable non-modal modes shown in
Fig.~\ref{fig4} must be higher with respect to the self-heating rate
for parametrically unstable waves.

\begin{figure}
  \resizebox{\hsize}{!}{\includegraphics[angle=0]{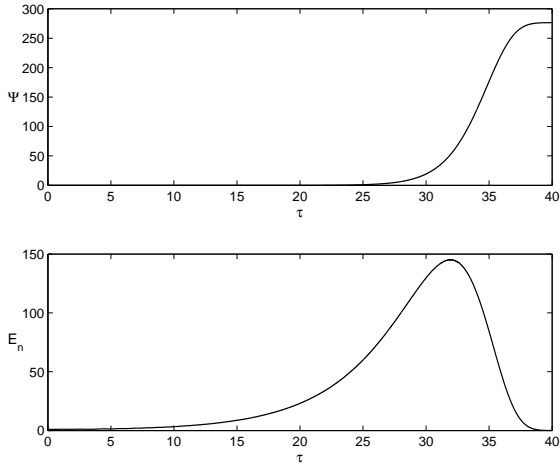}}
  \caption{The temporal behavior of $\Psi(\tau)$ and $E_n(\tau)$.
  The considered set of parameters is $\sigma = 0$, $a_1 = a_2 = 1$,
  $r_1 = r_2 = 0$, $\epsilon = 0.1$, $k_{x0}= 1$, $k_{y0}=
  0$, $u_{x0} = 0$, $u_{y0} = u_{z0} = 0$, $h_{x0} = h_{y0} = 0$,
  $d_0 = 0$, $\overline{\nu} = 1\times 10^{-5}$, $\overline{\eta} = 0$.}\label{fig5}
\end{figure}

In Fig.~\ref{fig5}, we show the temporal evolution of $\Psi(\tau)$
and $E_n(\tau)$. The considered set of parameters is the same as in
the previous figure, except for $\overline{\nu} = 1\times 10^{-5}$.
It is clear from the plots, that initially the instability
rate exceeds that of the self-heating (negligible role of the
dissipative factors), and therefore, the perturbation energy
increases. In the course of time, due to the exponential growth of
the wave vectors' components, the effective length scale,
$\lambda\sim 1/k$ decreases, leading to an increasing role of the
viscous terms. As a result, the waves become getting damped, the
perturbation energy reaches its maximum value, decreases and finally
vanishes (see Fig.~\ref{fig5}). This in turn, leads to a
self-heating rate with a saturated value, $\sim 270-280$, which is much
higher than $\Psi_{\infty}$ for the previous cases.

We study compressible MHD flows and all results are obtained for
$\epsilon = 0.1$. We did not discuss other regimes (like
$\epsilon\sim 1, \epsilon\gg 1$), because as it turned out, there is
no qualitative difference between the results obtained for this and
those regimes.

%%%%%%%%%%%%%%%%%%%%%%%%%%%%%%%%%%%%%%%%%%%%%%%%%%%%%%%%%%%%%%%%%%%%%%%%%%%%%
\section{Conclusions}
%%%%%%%%%%%%%%%%%%%%%%%%%%%%%%%%%%%%%%%%%%%%%%%%%%%%%%%%%%%%%%%%%%%%%%%%%%%%%%%

We have considered velocity shear induced instabilities in viscous
and resistive, kinematically complex MHD flows. By linearizing the
system of equations, two different kinds of instabilities have been
disclosed and studied: (a) the instability related to exponentially
evolving wave number vectors; and (b) the parametric instability
related to wave number vectors with periodic time dependence. The
latter instability works only in a relatively narrow area of
parametric space. In particular, it has been found that in 3D shear
flows one can find specific ranges of parameters, where the system
undergoes a very unstable regime (see Fig.~\ref{fig1}). By
considering the dissipative terms, it has been shown that for proper
values of viscosity and magnetic resistivity the self-heating rate
may be of the order of $10$ (see Fig.~\ref{fig2}). The self-heating
mechanism has also been studied former class of instabilities and it
was found out that in this case the efficiency of self-heating is
considerably higher than for the case of parametrically unstable
modes (see Fig.~\ref{fig5}). This is directly related to the fact
that in the time interval from the initial appearance of these modes
until the moment, when the dissipative terms become important,
exponentially evolving modes manage to amplify more efficiently than
periodically evolving (parametric) modes (see Fig.~\ref{fig4}).
Consequently, the energy given back to the background flow in the
form of heat is larger in the former case and the resulting
self-heating is more efficient too.

The present results show high efficiency of non-modal self-heating
in kinematically complex MHD shear flows, which may be of astrophysical
relevance for different classes of astronomical objects. In
particular, the solar atmosphere hosts various kinds of jet-like
structures possibly characterized by helical motion and swirling
velocity patterns \cite{pm98}. On the other hand, it is worth noting
that in the solar atmosphere there is the long-standing problem of
chromospheric and coronal heating \cite{bier,ruder}. A similar
problem of unclear origin of heating source exists in the young
stellar objects, particularly in Herbig-Haro jets. Therefore, it is
of great importance to check whether the non-modal self-heating
could be one of the mechanisms of plasma heating in these
astrophysical situations. Obviously it remains to see on a
quantitative level if the self-heating in MHD flows with kinematic
complexity may account for the actually observed heating in
concrete, realistic cases of astrophysical interest. Similarly, the results
obtained in this paper can be relevamt and important  for terrestrial (laboratory)
plasma MHD dissipative and resistive flows where the presence of velocity shear and kinematic
complexity is known.

%%%%%%%%%%%%%%%%%%%%%%%%%%%%%%%%%%%%%%%%%%%%%%%%%%%%%%%%%%%%%%%%%%%%%%%%%%%%%%%%
\section*{Acknowledgments}
%%%%%%%%%%%%%%%%%%%%%%%%%%%%%%%%%%%%%%%%%%%%%%%%%%%%%%%%%%%%%%%%%%%%%%%%%%%%%%%%%%%%%

The research of AR and ZO was supported by the Georgian National
Science Foundation grant GNSF/ST07/4-193. AR also acknowledges
partial financial support by the BELSPO grant, making possible his
visits to CPA/K.U.Leuven in 2010 and 2011. ZO acknowledges
hospitality of Katholieke Universiteit Leuven during his short term
visit in 2010. These results were obtained in the framework of the
projects GOA/2009-009 (K.U.Leuven), G.0729.11 (FWO-Vlaanderen) and
C~90347 (ESA Prodex 9).


\begin{thebibliography}{99}

\bibitem{tavec} Tavecchio et al.,
1993, ApJ, 614, 64
\bibitem{broder} Broderick, Avery E., Loeb, Abraham,
2009, ApJ, 703, 104L
\bibitem{kharb}
Kharb, P., Gabuzda, D. C., O'Dea, C. P., Shastri, P. \& Baum, S. A.,
2009, ApJ, 694, 1485
\bibitem{crabjet} Martin C. Weisskopf, Martin C.,
2000, ApJ, 536, 81L
\bibitem{pjet1} Johnson, S. P. \& Wang, Q. D., 2010, MNRAS, DOI:
10.1111/j.1365-2966.2010.17200.x
\bibitem{pm98} Pike C.D., Mason
H.E., 1998, Sol. Phys., 182, 333
\bibitem{tsun} Tsuneta, S., et al., 1991, Sol. Phys., 136, 37
\bibitem{shimo} Shimojo, M., \&
Shibata, K., 2000, ApJ, 542, 1100
\bibitem{sjet} Niita Nariaki V. et al.,
2008, ApJ, 675, L125
\bibitem{golub} Golub, L. et al., 2007, Sol. Phys., 243,
63
\bibitem{davis} Davis, C. J., Berndsen, A., Smith, M. D.,
Chrysostomou, A., \& Hobson, J., 2000, MNRAS, 314, 241
\bibitem{bacc} Bacciotti, F., Ray, T. P., Mundt, R., Eisloffel, J., \& Solf, J.
2002, ApJ, 576, 222
\bibitem{tref} Trefethen L.N., Trefethen A.E., Reddy S.C. Driscoll T.A., 1993,
Sience, 261, 578
\bibitem{rmb97} Rogava A.D., Mahajan S.M. Berezhiani V.I.,,
1997, Phys. Plasmas, 12, 4201
\bibitem{bod01b} Bodo G., Poedts S., Rogava A., Rossi
P., 2001, A\&A, 374, 337
\bibitem{andro} Rogava A.D., Mahajan S.M., Bodo G., Massaglia S., 2003, A\&A, 399,
421
\bibitem{chven} Rogava A.D., Bodo G., Massaglia S., Osmanov, Z., 2003, A\&A, 408,
401
\bibitem{androSH} Rogava
A. D., 2004, Ap. Space Sci., 293, 189
\bibitem{lcl06} Li J.W., Chen Y., Li Z.Y., 2006, Phys. Plasmas, 13, 042101
\bibitem{spp06} Shergelashvili B.M., Poedts S., Pataraya A.D.,
2006, ApJ, 642, L73
\bibitem{rop10} Rogava A.D., Osmanov Z., Poedts S., 2010, MNRAS, 404,
224
\bibitem{carroll} Carroll, Bradley, W. \& Ostlie, Dale, A., 2008,
An Introduction to Modern AStrophysics, Pearson
\bibitem{mahand} Mahajan S.M., Rogava A.D., 1999, ApJ,
518, 814
\bibitem{rog00} Rogava A. D., Poedts S.,
Mahajan S.M., 2000, A\&A, 354, 749
\bibitem{bier} Biermann L., 1946, Naturwissenschaften, 33, 118
\bibitem{ruder} Ruderman M. S.,
Nakariakov V. M., \& Roberts B. 1998, A\&A, 338, 1118
%\bibitem{kamio} Kamio,
%S. et al., 2010, A\&A, 510, L1

\end{thebibliography}
\end{document}